\documentclass[11pt,a4paper]{article}
\usepackage{epsfig}

\newlength{\dinwidth}
\newlength{\dinmargin}
\def\eq#1{{Eq.~(\ref{#1})}}
\newcommand{\Le}{\left(}
\newcommand{\Ra}{\right)}

\newcommand{\beq}{\begin{equation}}
\newcommand{\eeq}{\end{equation}}
\newcommand{\beqar}{\begin{eqnarray}}
\newcommand{\eeqar}{\end{eqnarray}}
\begin{document}

\title {{~}\\
{\Large \bf Conformal intercept of BFKL pomeron with NLO
running coupling constant corrections }\\}
\,
\author{ {~}\\
{~}\\
{\bf S.Bondarenko\,\,${}^{a)}$\,
\thanks{Email: sergey@fpaxp1.usc.es}}
\\[10mm]
{\it\normalsize ${}^{a)}$ University of Santiago de Compostela,}\\
{\it\normalsize Santiago de Compostela, Spain}\\}
\maketitle
\thispagestyle{empty}

\begin{abstract}
In the present note we propose a shift 
of the anomalous dimension function 
of the eigenfunctions of the BFKL equation 
with the NLO running coupling corrections.
The calculated eigenvalue of the modified equation turns out to be 
conformal invariant and we discuss consequences of this result.
\end{abstract}

\section{Introduction}

 The approaches based on the BFKL equation, \cite{Lip1,Lip2}, 
with the running coupling constant corrections included, \cite{Lip3,Camici1},  were 
considered initially a long time ago.
Some of these approaches were based on the
including of the running coupling effects via the 
simple redefinition of the coupling constant
into the running coupling constant in front of the equation,
\cite{Lip4,Lip5,Levin2,Nikolaev1,Kov3,Armesto1,Bondarenko1}, 
some were based on the more complicated "bootstrap" 
ideas applied to the BFKL kernel, \cite{Braun1,Levin1}. The new direct calculations
of the NLO kernel,  show a coincide of the "bootstrap"
approach with the results of diagram calculations, see \cite{Kov2,Kov1,Bal1},
in the quark sector of the corrections, 
i.e. in the number of flavors $n_{fl}$ leading order of the corrections. 
Still, whereas the "bootstrap" calculations include also a NLO
gluonic part of the kernel the direct calculation of these contributions were
completed only recently, see \cite{Fadin3,Fadin2,Fadin1,Bal2}. In the present note
we base our derivation on the base of "bootstrap" approach of the 
\cite{Braun1} and we do not consider a problem of the correctness of the obtained
gluonic part of NLO BFKL kernel.

 The common result of the considered earlier and present
calculations is that the eigenvalue of the NLO BFKL kernel
obtained with the use of the LO eigenfunctions is not 
conformal invariant anymore.
The so called running coupling
correction in the eigenvalue breaks the conformal invariance
of the equation and, therefore, makes impossible to consider
the NLO BFKL operator equation as the  equation with properly
found eigenfunctions and eigenvalue.
Therefore, due the fact that the used eigenfunctions
are not really eigenfunctions of the equation and that
found eigenvalue is not really eigenvalue of the operator
equation we face a problem of construction of the
Green's function of the equation, see \cite{Armesto1} for example.
The NLO Green's function, therefore, could be calculated
only approximately with the use of different perturbative schemes
in such a situation. 

 In the present note we propose a way to avoid this
difficulty. We propose to modify the equation by
including into the equation NLO corrections which arise from the
perturbative expansion of the eigenfunctions. Practically
it means that the anomalous dimension function of the eigenfunctions
in the NLO approximation is different from the LO anomalous 
dimension function. Perturbative expansion of the NLO anomalous
dimension function results in the redefinition of the eigenfunctions.
The following perturbative expansion of the eigenfunctions leads to the new
equation which has a conformal eigenvalue and, therefore, which could be used
for the construction of the NLO Green's function.

 The note is organized as follows. In the next section we remind 
LO and NLO results of the calculations of the BFKL operator equation, which
we will use in the further derivations. In the Section 3 we 
redefine a anomalous dimesion of the LO eigenfunctions and 
obtain a new equation with conformal invariant eigenvalue.
Section 4 dedicated to the Green's function of the 
new operator equation
and Section 5 is a conclusion of the note.

\section{LO and NLO BFKL kernel}

 First of all, let's remind, which kind of equation we consider.
The BFKL equation, \cite{Lip1,Lip2}, could be written as a operator equation
with the BFKL kernel and corresponding eigenfunctions and eigenvalues
\beq\label{BF1}
K\,\otimes\,\phi_{f}(k)\,=\,\omega_{f}\,\phi_{f}(k)\, 
\eeq
where $\phi_{f}$ is a eigenfunction and $\omega_{f}$ is
a eigenvalue of the equation with kernel $K$, see \cite{Lip2,Lip4}.
We begin from the LO BFKL kernel, \cite{Lip1,Lip2}, and consider the equation
at zero transferred momenta
\beq\label{BF2}
\alpha_{s}\,K_{LO}\,\otimes\,\phi_{\gamma}(k)\,=\,\frac{N_c\,\alpha_{s}}{\pi^2}\,
\int\,d^{2}\,\kappa\,\biggl[\frac{2}{\Le\,\kappa\,-\,k\,\Ra^{2}}\,
\phi_{\gamma}(\kappa)\,-
\,\frac{k^{2}}{\kappa^{2}\,\Le\,\kappa\,-\,k\,\Ra^{2}}\,\phi_{\gamma}(k)\biggr]\,
\eeq
where all momenta $k$ and $\kappa$ must be understand as a two dimensional vectors.
Due the non importance of that for the further derivations we do not underline
this fact especially, introducing the vector notation only where it will be need.
The eigenfunctions $\phi_{f}$ as a eigenfunctions of the operator equation 
must satisfy the completness relations
\beq\label{BF3}
\sum_{\gamma}\,\phi_{\gamma}(k)\,\phi_{\gamma}(\kappa)\,=\,
\delta^{2}\,\Le\,k\,-\,\kappa\,\Ra
\eeq
and must be orthogonal each to other 
\beq\label{BF4}
\int\,d^{2}\,k\,\phi_{\gamma}(k)\,\phi_{\gamma^{'}}(k)\,=\,
\delta\,(\gamma\,-\,\gamma^{'})
\eeq
The form of these eigenfunction is well known, it is
\beq\label{BF5}
\phi_{\gamma}(k)\,\propto\,\Le\,\frac{k^2}{\mu^2}\,\Ra^{\gamma}
\eeq
Here we omitted a normalization factor in front of the function and we
changed a usual definition of the eigenfunction introducing
some external scale $\mu$ into expression. It is clear, that this scale
is cancelled in the usual LO BFKL equation, leading only to the
redefinition of the Green's function of the equation. Defined
with the new eigenfunctions the Green's function will be dimensionless
instead the $k^{-2}$ dimension of the Green's function defined 
with the use of usual $\,k^{\gamma}$ eigenfunctions. 
The eigenvalue of the equation,
\beq\label{BF6}
\omega_{\gamma}^{LO}\,=\frac{N_{c}\,\alpha_{s}}{\pi}\,\Le\,
\,2\,\psi(1)\,-\,\psi(-\gamma)\,-\,\psi(1+\gamma)\,\Ra\,=\,
\frac{N_{c}\,\alpha_{s}}{\pi}\,\chi(-\gamma)
\eeq
calculated with the help of this eigenfunctions,
is the LO intercept of the BFKL Pomeron. 

 The NLO correction of the kernel, 
which are arising due the running coupling effect, were established
a long time ago on the basis of the bootsrap conditions 
applied to the kernel, see \cite{Braun1,Levin1}. The rule,
found for the introduction of the corrections in the \cite{Braun1}, is very simple.
Instead the LO propagator the following
propagator must be used in BFKL equation at NLO
\beq\label{BF7}
\frac{\alpha_{s}}{k^2}\,\rightarrow\,\frac{\alpha_{s}(k^2)}{k^2}\,=\,
\frac{\alpha_{s}}{1\,+\,\frac{\beta_{0}\,\alpha_{s}}{4\,\pi}
\,ln(k^{2}/\mu^{2})}\frac{1}{k^2}\,=\,\alpha_{s}\Le\,
\frac{1}{k^2}\,-\,\frac{\beta_{0}\,\alpha_{s}}{4\,\pi}
\,ln(k^{2}/\mu^{2})\frac{1}{k^2}\,\Ra\,
\eeq
where as usual $\beta_{0}\,=\,\frac{11\,N_{c}}{3}\,-\,\frac{2}{3}\,n_{fl}$\,
and as a renormalization scale we took the same $\mu^2$ as in 
Eq.\ref{BF5}
So, the NLO running coupling corrections determine
the following form NLO kernel
\beq\label{BF8}
\alpha_{s}^{2}\,K_{NLO}\otimes\phi_{\gamma}(k)=
-\frac{N_{c}\alpha_{s}^{2}\beta_{0}}{4\pi^3}
\int\,d^{2}\kappa\biggl[\frac{2}{\Le\kappa-k\Ra^{2}}
ln\Le\frac{\Le\kappa-k\Ra^{2}}{\mu^{2}}\Ra
\phi_{\gamma}(\kappa)-
\frac{k^{2}}{\kappa^{2}\Le\kappa-k\Ra^{2}}
ln\Le\frac{\kappa^{2}\Le\kappa-k\Ra^{2}}{k^{2}\mu^{2}}\Ra
\phi_{\gamma}(k)\biggr]\,
\eeq
We see, that obtained expression is coincide with the direct calculations
of $\cite{Kov1,Bal1}$ for the leading $n_{fl}$ order.
Now, using  Eq.\ref{BF5} eigenfunctions and methods of the calculation
of $\cite{Mueller1,Kov1}$ we have
\beq\label{BF9}
\frac{N_{c}\alpha_{s}^{2}\beta_{0}}{4\pi^3}
\int\,d^{2}\kappa\,\biggl[\frac{1}{\Le\kappa-k\Ra^{2}}
ln\Le\frac{\Le\kappa-k\Ra^{2}}{\mu^{2}}\Ra
\Le\frac{\kappa}{\mu}\Ra^{2\gamma}-
\frac{k^{2}}{\kappa^{2}\Le\kappa-k\Ra^{2}}
ln\Le\frac{\kappa^{2}\Le\kappa-k\Ra^{2}}{k^{2}\mu^{2}}\Ra
\Le\frac{k}{\mu}\Ra^{2\gamma}\biggr]\,=\,
\eeq
\[\,=\,
\frac{N_{c}\,\alpha_{s}^{2}\,\beta_{0}}{2\,\pi^2}\,
\biggl[2\,\psi^{2}(1)+2\,\psi(1)\,ln\Le\frac{k^2}{\mu^2}\Ra\,+\,
2\psi(1)\frac{\Gamma(1+\gamma)}{\Gamma(-\gamma)}
\frac{\partial}{\partial\,\gamma}
\frac{\Gamma(-\gamma)}{\Gamma(1+\gamma)}\,
\]
\[
\,+\,
ln\Le\frac{k^2}{\mu^2}\Ra\,\frac{\Gamma(1+\gamma)}{\Gamma(-\gamma)}
\frac{\partial}{\partial\,\gamma}
\frac{\Gamma(-\gamma)}{\Gamma(1+\gamma)}\,+\,
\frac{1}{2}\frac{\Gamma(1+\gamma)}{\Gamma(-\gamma)}
\frac{\partial^2}{\partial\,\gamma^2}
\frac{\Gamma(-\gamma)}{\Gamma(1+\gamma)}\biggr]\,
\Le\frac{k}{\mu}\Ra^{2\gamma}\,=\,
\]
\[
\,=\,\frac{N_{c}\,\alpha_{s}^{2}\,\beta_{0}}{2\,\pi^2}\Le\,
ln\Le\frac{k^2}{\mu^2}\Ra\,\chi(-\gamma)\,+\,
\frac{1}{2}\chi^{2}(-\gamma)
\,+\,\frac{1}{2}\,\psi^{'}(-\gamma)\,-\frac{1}{2}\,\psi^{'}(1+\gamma)\,\Ra\,
\Le\frac{k}{\mu}\Ra^{2\gamma}\,
\]
Summing up all terms together we obtain
\beq\label{BF10}
\alpha_{s}\,K_{LO}\,\otimes\,\phi_{\gamma}(k)\,+\,
\alpha_{s}^{2}\,K_{NLO}\otimes\,\phi_{\gamma}(k)=
\frac{N_{c}\,\alpha_{s}}{\pi}\,\biggl[\,
\chi(-\gamma)\,
\Le\,1\,-\,\frac{\alpha_{s}\,\beta_{0}}{2\,\pi}\,
ln\Le\frac{k^2}{\mu^2}\Ra\,\Ra\,-\,
\eeq
\[
\,-\,\frac{\alpha_{s}\,\beta_{0}}{2\,\pi}\,\Le
\,\frac{1}{2}\chi^{2}(-\gamma)
\,+\,\frac{1}{2}\,\psi^{'}(-\gamma)\,-\frac{1}{2}\,\psi^{'}(1+\gamma)\,\Ra\biggr]
\,\phi_{\gamma}(k)
\]
All these results are well known, see \cite{Kov1} for example. 
The only reason to reproduce these calculations
is the expression Eq.\ref{BF10}. Clearely, in spite of the 
Eq.\ref{BF6} the expression in the brackets in the r.h.s. of Eq.\ref{BF9}
is not eigenvalue of the BFKL equation. The $ln\Le\frac{k^2}{\mu^2}\Ra$
term, breaking conformal invariance of the expression,
makes impossible to interpetate the expression as the eigenvalue and, correspondingly,
as the intercept of the BFKL Pomeron. We see, that   
the functions Eq.\ref{BF5} are not eigenfunctions of the NLO BFKL kernel.

\section{Conformal intercept}

 Now we come back to the Eq.\ref{BF10} 
and will shift the anomalous dimension of the eigenfunction in this
expression
\beq\label{Eig13}
\gamma\,\rightarrow\,\gamma\,+\,\alpha_{s}\,\gamma_{1}\,=\,f
\eeq
In this case we have instead Eq.\ref{BF10}
\beq\label{Eig14}
\,\Le\,K\,\otimes\,\phi_{f}(k)\,=
\,\alpha_{s}\,K_{LO}\,\otimes\,+\,
\alpha_{s}^{2}\,K_{NLO}\,\Ra\,\otimes\,
\phi_{\gamma\,+\,\alpha_{s}\,\gamma_{1}\,}(k)\,=\,
\eeq
\[
\,=\,
\frac{N_{c}\,\alpha_{s}}{\pi}\,\biggl[\,
\chi(-\gamma\,-\,\alpha_{s}\,\gamma_{1}\,)\,
\Le\,1\,-\,\frac{\alpha_{s}\,\beta_{0}}{2\,\pi}\,
ln\Le\frac{k^2}{\mu^2}\Ra\,\Ra\,-\,
\]
\[
\,-\,\frac{\alpha_{s}\,\beta_{0}}{2\,\pi}\,\Le
\,\frac{1}{2}\chi^{2}(-\gamma)
\,+\,\frac{1}{2}\,\psi^{'}(-\gamma)\,-\frac{1}{2}\,\psi^{'}(1+\gamma)\,\Ra\biggr]
\,\phi_{\gamma\,+\,\alpha_{s}\,\gamma_{1}\,}(k)
\]
where we cared only about $\alpha_{s}^{2}$ order terms.
In order to continue further derivation let's assume that
the following expansion of the anomalous dimension function of the
eigenfunction holds
\beq\label{Eig3}
f=\,\gamma\,+\,\sum_{m=1}\,\alpha_{s}^{m}\,\gamma_{m}
\eeq
and, therefore, the perturbative expansion of the 
$\phi_{f}(k)\,$
functions over the complete set of initial
$\phi_{\gamma\,}(k)\,$ eigenfunctions in this case will have the following
form
\beq\label{Eig16}
\phi_{f}(k)\,=\,
\Le\frac{k^2}{\mu^2}\,\Ra^{f}\,=
\sum_{n=1}\,\alpha_{s}^{n-1}\,\gamma_{n-1}\,
\Le\,ln\Le\frac{k^2}{\mu^2}\Ra\,\Ra^{n-1}\,
\Le\frac{k^2}{\mu^2}\,\Ra^{\gamma}\,
\eeq
with $\gamma_{0}=1\,$.
Keeping in this expansion only
$\alpha_{s}$ order terms and incerting it 
back into the  \eq{Eig14} we obtain
\beq\label{Eig7}
K\,\otimes\,\phi_{f}(k)\,=\,\biggl[\,\alpha_{s}\,K_{LO}\,+\,
\alpha_{s}^{2}\,K_{NLO}\,\biggr]\,\otimes
\Le\,
1\,+\,\alpha_{s}\,\gamma_{1}\,ln\Le\,\frac{k^2}{\mu^2}\,\Ra\,\Ra
\Le\,\frac{k^2}{\mu^2}\,\Ra^{\,\gamma\,}\,=\,
\eeq
\[\,=\,
\,\biggl[\,\alpha_{s}\,K_{LO}\,+\,
\alpha_{s}^{2}\,\tilde{K}_{NLO}\,\biggr]\,\otimes
\Le\,\frac{k^2}{\mu^2}\,\Ra^{\,\gamma\,}\,=\,
\frac{N_{c}\,\alpha_{s}}{\pi}\,\biggl[\,
\chi(-\gamma\,-\,\alpha_{s}\,\gamma_{1}\,)\,
\Le\,1\,-\,\frac{\alpha_{s}\,\beta_{0}}{2\,\pi}\,
ln\Le\frac{k^2}{\mu^2}\Ra\,\Ra\,-\,
\]
\[
\,-\,\frac{\alpha_{s}\,\beta_{0}}{2\,\pi}\,\Le
\,\frac{1}{2}\chi^{2}(-\gamma)
\,+\,\frac{1}{2}\,\psi^{'}(-\gamma)\,-\frac{1}{2}\,\psi^{'}(1+\gamma)\,\Ra\biggr]
\,\phi_{\gamma\,+\,\alpha_{s}\,\gamma_{1}\,}(k)\,=\,
\]
\[
\,=\,
\frac{N_{c}\,\alpha_{s}}{\pi}\,\biggl[\,
\chi(-\gamma)\,-\,\frac{\alpha_{s}\,\beta_{0}}{2\,\pi}\,\Le
\,\frac{1}{2}\chi^{2}(-\gamma)
\,-\,\frac{1}{2}\,\psi^{'}(-\gamma)\,+\frac{1}{2}\,\psi^{'}(1+\gamma)\,\Ra\biggr]
\,\phi_{\,\gamma\,}(k)\,
\]
where we used
\beq\label{Eig10} 
\gamma_{1}\,=\,\frac{\beta_{0}}{4\,\pi}
\eeq
and  where we modified $\,K_{NLO}\,$ kernel adding to it
corrections which arise from $ln\Le\,\frac{k^2}{\mu^2}\,\Ra\,$
correction of the eigenfunction
\beq\label{Eig99}
\,\tilde{K}_{NLO}\,\otimes
\Le\,\frac{k^2}{\mu^2}\,\Ra^{\,\gamma\,}\,=\, 
-\frac{N_{c}\alpha_{s}^{2}\beta_{0}}{4\pi^3}
\int\,d^{2}\kappa\biggl[\frac{2}{\Le\kappa-k\Ra^{2}}
ln\Le\frac{\Le\kappa-k\Ra^{2}}{\kappa^{2}}\Ra
\phi_{\gamma}(\kappa)-
\frac{k^{2}}{\kappa^{2}\Le\kappa-k\Ra^{2}}
ln\Le\frac{\kappa^{2}\Le\kappa-k\Ra^{2}}{k^{4}}\Ra
\phi_{\gamma}(k)\biggr]\,
\eeq
So,  using  \eq{Eig13} shift
of the anomalous dimension of the eigenfunction,
we obtain for the NLO BFKL equation 
\beq\label{Eig11}
\,\biggl[\,\alpha_{s}\,K_{LO}\,+\,\alpha_{s}^{2}\,
\tilde{K}_{NLO}\,\biggr]
\,\otimes\,
\Le\,\frac{k^2}{\mu^2}\,\Ra^{\,\gamma\,}\,=\,
\Le\,\omega_{\gamma}^{LO}\,+\,\omega_{\gamma}^{NLO}\,\Ra
\,\Le\,\frac{k^2}{\mu^2}\,\Ra^{\,\gamma\,}\,
\eeq
where
\beq\label{Eig12}
\omega_{\gamma}^{NLO}\,=\,-\,\frac{N_{c}\alpha_{s}^2\,\beta_{0}}{2\,\pi^2}\,
\Le
\,\frac{1}{2}\chi^{2}(-\gamma)
\,-\,\frac{1}{2}\,\psi^{'}(-\gamma)\,+\frac{1}{2}\,\psi^{'}(1+\gamma)\,\Ra
\eeq
is the NLO conformal intercept of BFKL equation. 

 Definitely the same answer we obtain if we will  
calculate the correction to the $K_{NLO}$ kernel.
Calculating the following  
integral 
\beq\label{Eig8} 
\alpha_{s}^{2}\gamma_{1}K_{LO}\otimes
ln\Le\frac{k^2}{\mu^2}\Ra
\Le\frac{k}{\mu}\Ra^{2\gamma}=
\frac{N_c\,\alpha_{s}^{2}\,\gamma_{1}}{\pi^2}
\int\,d^{2}\kappa\biggl[\frac{1}{\Le\kappa-k\Ra^{2}}
ln\Le\frac{\kappa^{2}}{\mu^{2}}\Ra
\Le\frac{\kappa}{\mu}\Ra^{2\gamma}-
\frac{k^{2}}{\kappa^{2}\Le\kappa-k\Ra^{2}}
ln\Le\frac{k^{2}}{\mu^{2}}\Ra
\Le\frac{k}{\mu}\Ra^{2\gamma}\biggr]
\eeq
which gives
\beq\label{Eig9} 
\frac{N_c\,\alpha_{s}^{2}\,\gamma_{1}}{\pi^2}\,
\int\,d^{2}\kappa\,\biggl[\frac{1}{\Le\kappa-k\Ra^{2}}
ln\Le\frac{\kappa^{2}}{\mu^{2}}\Ra
\Le\frac{\kappa}{\mu}\Ra^{2\gamma}-
\frac{k^{2}}{\kappa^{2}\Le\kappa-k\Ra^{2}}
ln\Le\frac{k^{2}}{\mu^{2}}\Ra
\Le\frac{k}{\mu}\Ra^{2\gamma}\biggr]\,=\,
\eeq
\[
\,=\,\frac{2\,N_c\,\alpha_{s}^{2}\,\gamma_{1}}{\pi}\,
\biggl[2\,\psi(1)\,ln\Le\frac{k^2}{\mu^2}\Ra\,+\,
ln\Le\frac{k^2}{\mu^2}\Ra\,\frac{\Gamma(1+\gamma)}{\Gamma(-\gamma)}\,
\frac{\partial}{\partial\,\gamma}\frac{\Gamma(-\gamma)}{\Gamma(1+\gamma)}\,+
\]
\[
\,+\,\psi(1)\,\frac{\Gamma(1+\gamma)}{\Gamma(-\gamma)}\,
\frac{\partial}{\partial\,\gamma}\frac{\Gamma(-\gamma)}{\Gamma(1+\gamma)}\,+\,
\Le\,\frac{\partial}{\partial\,\gamma}\frac{\Gamma(-\gamma)}{\Gamma(1+\gamma)}\,\Ra
\Le\,\frac{\partial}{\partial\,\gamma}\frac{\Gamma(1+\gamma)}{\Gamma(-\gamma)}\,\Ra\,+
\,\frac{\Gamma(1+\gamma)}{\Gamma(-\gamma)}\,
\frac{\partial^{2}}{\partial\,f^{2}}\frac{\Gamma(-\gamma)}{\Gamma(1+\gamma)}\,
\biggr]\,\Le\frac{k}{\mu}\Ra^{2\gamma}\,
\]
and adding this expression to the Eq.\ref{BF10} we again obtain 
Eq.\ref{Eig12} answer.

\section{Green's function of the NLO BFKL equation}

 Now we consider a solution of NLO BFKL equation, i.e.
Green's function of the equation constructed with the help of the 
found eigenfunctions. The Green's function of the \eq{Eig11} is
\beq\label{Gr1}
f(\omega,k_1,k_2)\,=\,\frac{1}{2\,\pi^2\,}
\int\,d\gamma\,
\Le\frac{k_{1}^{2}}{\mu^2}\Ra^{\gamma}\,
\Le\frac{k_{2}^{2}}{\mu^2}\Ra^{\gamma^{*}}\,
\frac{1}{\omega\,-\,\omega_{\gamma}^{LO}\,-\,\omega_{\gamma}^{NLO}}
\eeq
for the case of conformal spin $n=0$.
Coming back to the full anomalous dimension function
of the eigenfunction from \eq{Eig13}
\beq\label{Gr2}
\gamma\,=\,f\,-\,\frac{\alpha_{s}\,\beta_{0}}{4\,\pi}\,,
\eeq  
we obtain
\beq\label{Gr3}
f(\omega,k_1,k_2)\,=\,\frac{1}{2\,\pi^2\,}
\int\,d\,f\,
\Le\frac{k_{1}^{2}}{\mu^2}\Ra^{f-\,\frac{\alpha_{s}\,\beta_{0}}{4\,\pi}\,}\,
\Le\frac{k_{2}^{2}}{\mu^2}\Ra^{f^{*}-\,\frac{\alpha_{s}\,\beta_{0}}{4\,\pi}\,}\,
\frac{1}{\omega\,-\,\omega_{f}}
\eeq
where
\beq\label{Gr4}
\omega_{f}\,=\,
\frac{N_{c}\,\alpha_{s}}{\pi}\,\biggl[\,
\chi(-f)\,-\,
\frac{\alpha_{s}\,\beta_{0}}{2\,\pi}\,\Le
\,\frac{1}{2}\chi^{2}(-f)
\,+\,\frac{1}{2}\,\psi^{'}(-f)\,-\frac{1}{2}\,\psi^{'}(1+f)\,\Ra\biggr]
\,
\eeq
Redefining $f$ as
\beq\label{Gr5}
f\,=\,-1/2\,-\,i\,\nu\,
\eeq
we obtain our final expression for the Green's function of the equation
\beq\label{Gr6}
f(\omega,k_1,k_2)\,=\,\frac{\mu^2}{2\,\pi^2\,k_{1}\,k_{2}}
\int_{-\infty}^{\,\infty}\,d\,\nu\,
\Le\frac{k_{2}^{2}}{k_{1}^{2}}\Ra^{\,i\,\nu\,}\,
\frac{1}{\omega\,-\,\omega_{\nu}}\Le\,
1\,-\,\frac{\alpha_{s}\,\beta_{0}}{4\,\pi}\,ln\Le\,
\frac{k_{1}^{2}\,k_{2}^{2}\,}{\mu^4}\Ra\,
\,\Ra
\eeq
with
\beq\label{Gr7}
\omega_{\nu}\,=\,
\frac{N_{c}\,\alpha_{s}}{\pi}\,\biggl[\,
\chi(\nu\,)\,-\,
\frac{\alpha_{s}\,\beta_{0}}{2\,\pi}\,\Le
\,\frac{1}{2}\chi^{2}(\,\nu\,)
\,-\,\frac{1}{2\,i}\,\psi^{'}(1/2\,+\,i\,\nu\,)\,+
\frac{1}{2\,i}\,\psi^{'}(1/2\,-\,i\,\nu\,)\,\Ra\biggr]
\,
\eeq
The diffusion approximation for the Green's function we obtain
expanding $\omega_{\nu}\,$ over $\nu$
\beq\label{Gr8}
\omega_{\nu}\,=\,\omega_{0}\,-\,a^2\,\nu^{2}\,
\eeq
with
\beq\label{Gr9}
\,\omega_{0}\,=\,4\,\frac{N_{c}\,\alpha_{s}}{\pi}\,ln\,2\,
\biggl[\,1\,-\,
\frac{\alpha_{s}\,\beta_{0}}{\pi}\,\Le\,ln\,2\,-\,
\frac{\pi^2}{16}\,\Ra\,\biggr]\,
\eeq
and
\beq\label{Gr10}
\,a^2\,=\,14\,\frac{N_{c}\,\alpha_{s}}{\pi}\,\zeta\,(3)\
\Le\,1\,-\,
\,2\,\frac{\alpha_{s}\,\beta_{0}}{\pi}\,\,ln\,2\,\Ra
+\,\frac{N_{c}\alpha_{s}^2\,\beta_{0}\,\pi^2}{4}\,
\eeq
Integration of the expression 
over $\omega$ and $\nu$ variables gives  
final answer for the Green's function in the diffusion approximation
\beq\label{Gr11}
F(s,k_1,k_2)\,\approx\,
\frac{\mu^2}{2\,\pi\,a\,k_{1}\,k_{2}}
\Le\frac{s}{s_0}\Ra^{\omega_{0}}\,
\frac{1}{\sqrt{\pi\,ln\Le\,s/s_{0}\Ra}}\,
exp\,\Le\,-\frac{ln^2\,\Le\,k_{1}^{2}\,/\,k_{2}^{2}\,\Ra}
{4\,a^2\,ln\Le\,s/s_{0}\Ra}\,\Ra
\Le\,1-\,
\frac{\alpha_{s}\,\beta_{0}}{4\,\pi}\,ln\Le\,
\frac{k_{1}^{2}\,k_{2}^{2}\,}{\mu^4}\Ra\,\Ra
\eeq
with the $\omega_{0}$ and $a$  from \eq{Gr9} and \eq{Gr10} correspondingly.

\section{Conclusion}

 The shift of the  anomalous dimension function of the
LO eigenfunctions, represented by   \eq{Eig13},
is justified by the fact that in the NLO 
approximation the functions \eq{BF5}
are not eigenfunctions of the NLO equation. 
Therefore, transition from the 
\eq{BF10} to the \eq{Eig14} does not affect on the correctness of the
operator equation. Nevertheless, after a shift and redefinition of the
NLO BFKL kernel we obtain equation \eq{Eig11} which is correctly defined as a 
operator equation with proper eigenfunctions and correspondingly
conformal eigenvalue. From the formal point of view this shift
reflects the functional structure of the $\gamma$ in \eq{BF5}.
Indeed, being function of $\alpha_{s}$ we can assume, that in the NLO
approximation
the LO $\gamma$ acquires some corrections which are
clarified in the shift $\gamma\rightarrow\,f$. Physically it means that 
together with NLO corrections of the kernel we need to account the NLO corrections 
of the anomalous dimension function of the eigenfunctions
and this precisely that the shift \eq{Eig13} means. 
There is a strong assumption behind 
this statement, we assume that the functional
form of the eigenfunctions does not change when NLO 
corrections of the kernel are considered and 
that all NLO corrections of the eigenfunctions
are accumulated in the anomalous dimension function
of the eigenfunction. This proposition could not be prooven directly
and therefore we could consider the  \eq{Eig11} as a effectively
constructed operator equation where the request of the  conformal eigenvalue
determines a coefficient $\gamma_{1}$ in the \eq{Eig3} expansion.

 The expansion \eq{Eig3} we can interpretate also as a perturbative 
expansion of the "full" eigenfunction in the case when we know only a 
part of NLO corrections to the kernel. In this case the redefinition \eq{Eig16} of the "shifted" or "full" eigenfunctions
in the terms of initial "non-shifted"  
eigenfunctions looks like
a renormalization group transition from one basis of eigenfunctions to another.  
As a  results of this transition we have a new equation 
with the  
conformal eigenvalue and with the 
NLO corrections to the  Green's function of the equation.
Indeed, one of the main advantages of the proposed framework is the possibility
to construct a Green's function of the equation. Using
eigenfunctions of the equation the definition of the Green's function is standart and simple. Considering the Green's function of the equation,
see  \eq{Gr6}, we obtain that the shift \eq{Eig13} leads to the simple
form of the NLO corrections to the LO Green's function and 
that the form of these corrections is
depend on the coefficients $\gamma_{i}$ in the expansion \eq{Eig3}.
Another advantage of the 
proposed framework is also the conformal structure of the  eigenvalue which we
consider as NLO Pomeron intercept and which could be
written in the "diffusion" approximation, see \eq{Gr9} and \eq{Gr10}.

 Another interesting property of the proposed approach is that 
we considered dimensionless theory, introducing some external scale 
$\mu^2$ even in LO approximation. The influence of this fact 
on the final result and the relation 
of proposed framework with the result of the NLO BFKL equation 
with the broken  conformal invariance of the eigenvalue
is a interesting subject which we plane investigate in our future studies.

\section*{Acknowledgments}

\,\,\,S.B. especially grateful to N.Armesto, E.Levin and L.Lipatov for the 
usefull remarks about the subject of the note.
This work was done with the support of the Ministerio de Educacion
y Ciencia of Spain
under project FPA2005-01963 together with Xunta de Galicia
(Conselleria de Educacion).

%%%%%%%%%%%%%%%%%%%%%%%%%%%%%%%%%%%%%%%%%%%%%%%%%%%%%%%%%%%%%%%%%%%%%%%%%%%%

\end{document}